\documentclass[10pt,letterpaper]{article}
\usepackage{url,graphicx,cite,gensymb}
\pdfoutput=1

\title{\bf A Review of the Enviro-Net Project$^{*}$}

\begin{document}

\maketitle

\vspace{0.1cm}

\begin{center}
{\large
Gilberto Z. Pastorello$^{1}$\\
G. Arturo Sanchez-Azofeifa$^{1}$\\
Mario A. Nascimento$^{2}$\\
}

\vspace{0.6cm}

{\small
$^{1}$ Department of Earth and Atmospheric Sciences\\
1-26 Earth Sciences Buiding\\
University of Alberta\\
T6G 2E3\\
Edmonton, Alberta, Canada.\\
\url{gilbertozp@acm.org}, \url{arturo.sanchez@ualberta.ca}\\
}

\vspace{0.4cm}

{\small
$^{2}$ Department of Computing Science\\
2-32 Athabasca Hall\\
University of Alberta\\
T6G 2E8\\
Edmonton, Alberta, Canada.\\
\url{mario.nascimento@ualberta.ca}
}
\end{center}

\vspace{0.4cm}

{\noindent\scriptsize
$^{*}$Text published in:\\G.~Z.~Pastorello, G.~A.~Sanchez-Azofeifa, M.~A.~Nascimento. {\em Enviro-Net: From Networks of Ground-Based Sensor Systems to a Web Platform for Sensor Data Management}. {\bf Sensors}. 2011. 11(6):6454-6479. doi: 10.3390/s110606454
}

\vspace{0.4cm}

\abstract{ Ecosystems monitoring is essential to properly understand
their development and the effects of events, both climatological and
anthropological in nature. The amount of data used in these
assessments is increasing at very high rates. This is due to
increasing availability of sensing systems and the development of
new techniques to analyze sensor data. The Enviro-Net Project
encompasses several of such sensor system deployments across five
countries in the Americas. These deployments use a few different
ground-based sensor systems, installed at different heights
monitoring the conditions in tropical dry forests over long periods
of time. This paper presents our experience in deploying and
maintaining these systems, retrieving and pre-processing the data,
and describes the Web portal developed to help with data management,
visualization and analysis. }

\vspace{0.4cm}

\section{Introduction}
\label{intro}

Monitoring ecosystems at high spatial and temporal resolutions still
is a challenging endeavor. Satellite-embarked sensors that offer
regular passes support only coarse resolution monitoring and
on-demand high resolution satellite or airborne-based monitoring are
still too expensive to be considered viable options for frequent
data collections. Furthermore, validation of satellite and airborne
measurements against the values observed at ground level is often
difficult to obtain. Ground-based, or {\em in-situ}, sensor systems
for environmental monitoring have associated challenges as well
\cite{rugral2009}, but have undergone a considerable evolution
recently. Such systems are now capable of collecting data at very
high temporal resolution for very specific ecosystems through long
periods of time. In particular, the use of wireless sensor systems
has been shown to be very effective in this type of monitoring
\cite{poarbr2005}, from the cost perspective and increasingly in
terms of performance and reliability as well.

There are many challenges associated with high resolution (both
spatial and temporal) {\em in-situ} environmental monitoring, many
of which already well recognized in the literature. Rundel \emph{et
al.} \cite{rugral2009}, for instance, discuss how these networks
generate more data than can be managed by traditional methods for
field research data, with data quality assurance and control
surpassing capabilities of single individuals dealing with the data,
but still being required to produce high-quality data. The large
variety of problems impacting quality can be more easily detected by
using adequate cyberinfratructure for automating the detection,
which also allows more timely identification of problems in the
deployments themselves. They also argue that, although data storage
and retrieval is reasonably easy to attain, publishing and sharing
data is not as straightforward. Still according to the authors, one
of the advantages of this integrated approach for offering data from
multiple sensors is the larger world view generated, which is not
possible with single sensors---at least not at these spatio-temporal
scales. The authors also acknowledge the importance of training
scientists in using {\em in-situ} monitoring tools, the flexibility
of power requirements for these systems (especially wireless) and
the use of energy harvesting, problems related gaps in the data
(from numerous causes), difficulty to assess precision and fidelity
in such systems, and the value of commercial availability for
decreasing costs and scaling up deployments sizes.

Hart and Martinez \cite{hama2006} discuss power management, large
volumes of data and required cyberinfrastructure, beginning of
commercial efforts, and data quality control as important issues
concerning {\em in-situ} environmental monitoring. They also raise
additional points that require more work, such as assessment of
environmental conditions any equipment needs to withstand them (e.g.,
temperature, pressure, vibration); standardization requirements
related to data and metadata representation; security requirements,
preventing tampering with both equipment and datasets within the
data management systems; and, better means for data interpretation
(e.g., by using new methods for data mining). Another relevant
effort can be found in the report from Estrin \emph{et al.}
\cite{esmibo2003}, who focus on cyberinfrastructure. Key points
include: the need for better prototyping and design of end-to-end
test-beds to allow validation across wide ranges of environments,
applications and domains; creation of better services regarding time
synchronization, {\em in-situ} calibration, and adaptive duty
cycling, among others; seamless use of high performance computing
facilities for data processing; tools to improve support for
metadata; and, collaboration efforts as a basis for training new
scientists (from multiple domains) and as a mechanism for sustaining
long term deployments.

This paper distills our experience in deploying and managing {\em
in-situ} sensor systems within the Enviro-Net Project
(\url{http://www.enviro-net.info/}). Currently, Enviro-Net includes 39
deployments spread throughout nine sites in six different countries
(Argentina, Brazil, Canada, Costa Rica, Mexico and Panama), and is
coordinated at the University of Alberta, in cooperation with local
partner research teams at each site. The initial goal of the
deployments was to monitor vegetation phenology, the study of
climate effects on periodic biological activity \cite{shwartz2003},
correlating it with environmental variables, such as availability of
light, air temperature, \emph{etc.} These and other variables are
monitored by different types of sensing systems, with the collected
data being transmitted back to Internet servers in Alberta either
through a commercial satellite up-link or being manually retrieved
from the data loggers and then sent via email, FTP or Enviro-Net's
website. The following gives but one example of the applicability
and usefulness of such type of systems. From the data collected by a
combination of two types of specialized solar radiation fluxes
sensors, it is possible to derive different vegetation indexes,
which can be used as proxies to monitoring phenological responses.
In order to distinguish vegetation distribution, particularly from
perspectives such as species distribution or successional stage, the
areas to be monitored are numerous and relatively small. Similarly,
short term effects of isolated climatic phenomena (e.g., a rainstorm
or sharp changes in temperature) require higher rates of data
acquisition. These characteristics require higher spatial and
temporal resolutions only achieved through {\em in-situ} monitoring
of each ecosystem.

In this context, detailed discussions of how we dealt with these
challenges within the Enviro-Net Project form the main
contributions of this paper, particularly considering the scenario
under which the project was developed. The monitored sites are
mostly tropical dry forests in remote locations, which are
challenging environments for both equipment performance and
personnel's ability to work. Also, all deployments are based on
inexpensive and commercially available technology, essential
characteristics to allow scalability and reproducibility of
experiments. The heterogeneity of equipment from different
manufacturers also introduce difficulties, mainly regarding systems
maintenance. Having long term (multiple-year) deployments impose
extra management requirements. Integrated data management, a fourth
aspect, presents numerous challenges ranging from data quality
control to user interface usability. Finally, and maybe the most
relevant aspect, is the issue of high spatial and temporal
resolutions, considered not only within a single deployment, but
also among different deployments both in the same and different
sites. Part of these challenges have simple individual solutions,
however, from a more holistic perspective, the integration of the
solutions for all of them is what actually enables the use of
sensors systems for {\em in-situ} environmental monitoring. After a
review of related work on Section~\ref{related}, this paper
describes our solutions regarding deployments of {\em in-situ}
monitoring systems in Section~\ref{deployment}, pre-processing and
treatment of data in Section~\ref{ingestion} and data publication
and accessibility using a Web-based system in Section~\ref{web}.

\section{Related Work}
\label{related}

This section divides related work discussion into applications
(covering the motivation for {\em in-situ} monitoring), deployments
(showing experiences in installing and maintaining sensor systems),
and data management (comparing different efforts in dealing with the
large amounts of sensor data generated).

\subsection{Applications}
\label{app}

Environmental monitoring is one of the driving forces behind the
adoption of ground-based sensing systems, pushing the need for
higher spatial and temporal resolution. Examples of efforts in this
direction include: (i) the creation of the National Ecological
Observatory Network (NEON)  \cite{kescha2008}, which aims at
studying climate change, land-use change and invasive species on a
continental scale using, among other methods and technologies,
ground-based deployments of sensor systems; (ii) FLUXNET
\cite{bafagu2001}, which use micrometeorological and flux towers to
measure exchanges of carbon dioxide, water vapor, and energy between
terrestrial ecosystems and the atmosphere. These initiatives heavily
rely on long term ground-based monitoring solutions. FLUXNET has a
public data management framework called Fluxdata.org
\cite{huagin2009}, which also offers flexible metadata support.
However, due to the diversity of equipment and protocols for
deployment and data pre-processing, data integration within the
Fluxdata.org system is limited, mostly offering access to data on
the original format provided by the data producers. This limits the
possibilities of applying filters and aggregation operations to the
data or generating derived data products within the system. Although
our system also deals with a variety of equipment, the deployment
protocols are largely uniform, and pre-processing protocols are
developed using a centralized approach, which allow us to achieve a
considerable level of data integration within Enviro-Net.

These and other initiatives, aiming at integration of ground-based
monitoring efforts, are leading to an evolution from single site
environmental monitoring into networks for environment
observation~\cite{hama2006}. This evolution culminates with the
current vision for a {\em Sensor Web}
\cite{delin2002,teillet2010,bopere2008}, encompassing several types
of deployments of sensor systems, interconnecting them globally
through a Web-based integration strategy using standards developed
by the Sensor Web Enablement
(\url{http://www.opengeospatial.org/projects/groups/sensorweb}) Working
Group of the Open Geospatial Consortium, Inc. (OGC)
(\url{http://www.opengeospatial.org/}).

A small clarification on the definition for (wireless) sensor
networks may be in order. Mainly within Computing Science (CS)
research \cite{chku2003,mahaon2004} and in earlier Sensor Web
related efforts \cite{delin2002}, this definition is narrower than
what is used in this paper. In this more restrictive definition, a
(wireless) sensor network is based on nodes (also known as
``motes'') that have sensing, data storage/processing, and
communication components plus a power source. These nodes are
usually autonomous and operate cooperatively---by communicating
amongst themselves---to collect and process data, also being programmable,
 \emph{i.e.}, able to behave differently according
to, for instance, the type of application, power supply conditions,
environmental conditions, \emph{etc.} Although we have used this
type of wireless nodes in our deployments, we do not require the
capability of offering communication amongst network's components.
Instead, we adopt the centralized type of processing architecture as
classified by \cite{chku2003}, being more in line with the current
Sensor Web approach to networks \cite{bopere2008}. It is sufficient
for us, for instance, that the connection of sensing elements be
done at the level of integrated data products.

Applications of {\em in-situ} monitoring systems are also the topic
of other research efforts. Porter \emph{et al.}
\cite{poarbr2005} present a good review of the capabilities of
wireless sensor networks (WSN) to be applied within the ecological
domain. Hamilton \emph{et al.} \cite{hagrru2007}, while covering
capabilities of networks of sensors applied to ecology as well,
also highlight the idea of ecological observatories, adopted
within NEON. An extensive review of {\em in-situ} monitoring
efforts is presented by Rundel \emph{et al.} \cite{rugral2009},
classified according to their area of focus: above ground,
under-ground, and aquatic environments. Porter~\emph{et
al.}~\cite{ponakr2009} discuss the state of the sensing technology,
what can already be accomplished and a few areas that require more
development (e.g., data management software and new types of
sensors). Precision agriculture is a particularly relevant
application area for pervasive sensing technology. For instance, Lee
\emph{et al.} \cite{lealya2010} evaluate monitoring applied to
specialty crop, while Matese \emph{et al.} \cite{mageza2009} use
wireless sensor network in vineyard monitoring, and Aquino-Santos
\emph{et al.} \cite{aqgoed2011} evaluate data transmission
protocols in small scale deployments in watermelon fields. In this
paper, we discuss aspects that apply to many of these scenarios,
particularly if considering them in a long term monitoring
perspective. However, our focus is on practical and logistics
aspects of deploying and maintaining equipment, retrieving and
managing the data, and supporting analysis of data products.

\subsection{Deployments}
\label{dep}

Other research groups have discussed their efforts with ground-based
deployments of sensor systems, mostly focusing on the use of
wireless equipment. A pioneering effort in applying wireless sensor
networks was the habitat monitoring experiment in the Great Duck
Island \cite{maposz2002} in the coast of Maine in the United States,
deployed to offer a less intrusive way to study behavior and nesting
of seabird colonies. The SensorScope project
\cite{bainsc2008a,inbasc2010} is another example, taking place
mainly in Switzerland. They have described their experience with
developing the hardware and software for their wireless system,
performing tests, and going on deployment expeditions, along with
their architecture and communication protocols. With a focus on
solar energy availability, AdaptSens \cite{wayano2009} adopts
system-wide levels of operation to cope with different amounts of
available energy. GreenOrbs
(\url{http://www.greenorbs.org/})~\cite{moheli2009} is a long term
effort for monitoring an university campus urban forest close to
Hangzhou in China, using a large number of nodes. LUSTER
\cite{sewoca2007} is a system for monitoring ecological variables
that implements fault-tolerant distributed storage over a
delay-tolerant network using an hierarchical architecture; the
system also covers user interaction both in the field expeditions
and a web interface for data retrieval. Another effort
\cite{miyado2008}, aiming at monitoring the UNESCO World Heritage
site Mogao Grottoes in Dunhuang, China, implemented a low power
wireless monitoring system inside the site's caves with a tailored
long distance connection to transmit the data back to an on-line
server. Another World Heritage site, a rainforest ecosystem in
Queensland, Australia, was monitored by a wireless sensor network
project \cite{wahuco2008}, which served as a prototype for future
long term deployments using similar configurations. Another
interesting application, monitoring the activities of volcanoes in
Ecuador \cite{welojo2006,welsh2010}, entails addressing
issues such as higher sampling rates (100 Hz or more), need for
higher accuracy and more expensive sensors. Changing the spatial
scale a little, monitoring a single redwood tree \cite{toposz2005}
in California in the United States, offered new insight in
understanding the microclimate surrounding this type of tree.
Reports on deployment experiences also focus on the diversity of
problems faced when using wireless sensing equipment, such as the
LOFAR-agro project~\cite{labavi2006} that experienced problems
ranging from hardware failures to network protocols errors and
software problems. While deployment related efforts in our work
focus on issues related to managing the life cycle of ground-based
sensor data, other works \cite{szpoma2004,jiar2009} bring
evaluations of technology for wireless sensor network equipment,
including communication protocols, power consumption and data
transmission~issues.

To the best of our knowledge, none of the deployment efforts
reviewed here address the same scenario as ours: having (multi-year)
long term deployments, based on cooperative efforts of several
(heterogeneous) teams, using commercially available equipment from
multiple manufacturers, with an integrated effort of data retrieval,
quality control and data availability through an easy to use
Web-based platform. We believe this is a more realistic scenario for
ground-based environmental monitoring efforts. The current efforts
within the Life Under Your Feet project
(\url{http://lifeunderyourfeet.org/}) \cite{mutesz2006} are the closest
to our own, also having long term, spatially distributed deployments
with a Web-based data visualization interface integrated with
geolocation information. However, they do not seem to deal with
heterogeneous equipment and data formats, nor offer
filtering/aggregation options, derived datasets or quality
information in their data management solution.

\subsection{Data Management}
\label{dataman}

Many of the challenges related to sensor data management have been
known for a while \cite{esmibo2003}. However, several technical and
non-technical questions still remain unaddressed. Broad scope
projects for management of earth observation data try to present a
top-down approach to data management. One such project is DataOne
(\url{http://www.dataone.org/}), an effort towards distributed
cyberinfrastructure for Earth observation data, bringing together a
multitude of data providers and consumers. Another effort is our
partner project GeoChronos (\url{http://www.geochronos.org/}), which
implements means for sharing (and interacting with) tools, datasets
and libraries of records within the Earth observation domain.
Enviro-Net, however, uses more of a bottom-up approach, offering
specialized solutions for the types of data supported, expanding
these types as needed. This allows data management solutions that
are geared towards specific needs to answer specific science
questions.

Although it is common to think about sensor data management as
stream data management, with the associated challenges (on-line
aggregation, classification, \emph{etc.}) \cite{olgr2008}, at least
within environmental research, particularly in ground-based
monitoring, this is not a frequent scenario. Most of the current
applications based on sensor data use the perspective of historical
(or an archive of) time series data. Applications using the stream
data perspective are only beginning to appear, and the current
applications that do require that perspective---e.g., volcano
monitoring \cite{welojo2006}---are still the exception. Data
manipulation for most of the current applications is done after
having the data collected and stored, applying a variety of
analytical operations in an offline fashion
\cite{ozgrsz2006,huagin2009}.

Middleware software for automating control of deployments is also
the focus of current research efforts, in form of architectures for
integrating different network deployments \cite{abhasa2007}, or
Web-based interfaces for interaction with and control of wireless
deployments \cite{stda2009}. Our focus, on the other hand, is on
managing the data products rather than controlling the equipment
from within our system.

The data archival aspect of data management involves not only
storage of data, but also retrieval, documentation, access control,
among other issues. Furthermore, data curation of long-term
repositories involves not only handling the data but also helping
scientists answering research questions and also maintaining the
underlying computational infrastructure \cite{kaba2008}. Within
Enviro-Net, although we are only beginning to to devise our long
term plans for infrastructure maintenance, our system already offers
data access with a number of flexibility aspects to foster efficient
use of the data. Efforts on applying digital library practices in
support of sensor data management are also gaining acceptance
\cite{bowama2007}. Issues of data quality and integrity, as well as
the elements of data collection that affect them, need to be an
integral part of such efforts \cite{waboma2007}, particularly from
the perspective of making data documentation available along with
the datasets. In this scenario, metadata becomes as valuable as the
datasets themselves, from quality metadata about deployments
\cite{lustgu2009}, to offering search and annotation options and
enriching visualization~\cite{dakumi2008}. Finally, Application
Programming Interfaces (APIs) allow data to be accessed in a
programmatic way, which can be achieved, for instance, using Web
services interfaces (using Sensor Web Enablement standards) or using
specialized solutions such as a wrapper-based middlewares
\cite{cagama2009} or REST-based APIs \cite{guudpo2010}. Data quality
aspects are an integral part of Enviro-Net, and are being improved,
particularly regarding documentation and metadata coverage. Although
data ingestion is largely automated and data access is possible
through the Web user interface within Enviro-Net, data access using
a programmatic interface is still under development.

\section{Sensor Systems Deployments}
\label{deployment}

Apart from a few test installations, all of our deployments are
intended to be long term, collecting data for a minimum of two to
three years. The earliest deployments were installed in mid 2007,
with the first wireless deployments installed in mid 2008. All
deployments suffered from interruption in data collection on some
scale, usually from a few days up to a couple of months, depending
on how early the problem was detected. Earlier deployments suffered
100\% failure rate due to equipment design being incompatible
with tropical environments. Many problems were related to unexpected
interactions of environmental conditions with the equipment.
However, most of the deployments are still operational today, with
secured funding for maintaining them operational until at least
2013.

Currently, Enviro-Net has 39 permanent deployments, plus temporary
deployments in Edmonton, Canada for equipment testing and
calibration. The {\em Biosphere Reserve of Chamela-Cuixmala} in the
state of Jalisco, Mexico has a tower (overlooking the top of the
canopy) and a wireless understory sensor system. The number of nodes
in a wireless deployment is usually 12, but there are deployments
with as few as five and as many as 20 nodes, each node having
between three to six sensors each. The {\em Santa Rosa National
Park} in Costa Rica hosts two more towers. The {\em Parque Natural
Metropolitano} in Panama has the most recent deployment with 24
thermocouples monitoring leaf temperatures. In Brazil, three sites
have deployments: the {\em Mata Seca State Park}, the {\em Serra do
Cip\'o National Park}, and the {\em Environmental Protection Area of
the Pandeiros River}, all located in the Minas Gerais state. The
Mata Seca park hosts five towers and eight understory deployments
(including four wireless deployments), all in the {\em cerrado}
ecosystem, which is similar to a savanna; three understory
deployments are active close to the Pandeiros river, also a {\em
cerrado} ecosystem; and, Serra do Cip\'o park has five towers plus
seven understory deployments, two of which using wireless systems,
covering natural grasslands and forest vegetation in the {\em
cerrado}. Finally, three deployments are operational in the province
of San Luis in Argentina, a phenology tower monitoring a grassland
ecosystem, and one tower and one wireless understory deployment
installed in a adjacent {\em chaco} ecosystem. Two more wireless
towers are operational {\em chaco} and grassland ecosystems in the
province of C\'ordoba, Argentina. Three more deployments are expected
to start data collection in 2011 in the province of San Luis. Although
the ideal spatial scales for many applications requires higher
numbers of nodes deployed to be considered high spatial
density---more in line with our plans for future sensor
networks---the intermediary step with 5--20 nodes per deployment was
necessary to prove this kind of scale is feasible in remote
locations with long term deployments. These are, however, dense
enough to characterize many ecosystem level behavior (such as
response to climatic events), and even differences between
neighboring ecosystems. The experience acquired in these smaller
deployments, which is the fundamental contribution of this text,
serves as a basis for these larger scales expeditions.

The main challenge of having deployments across an entire continent
is without question maintaining them. Partnerships with research
groups based closer to the deployment sites proved essential, with
the added issue of offering training to the people performing basic
maintenance. The small amounts of time available for training leads
to the choice of equipment that is simple to use and maintain. Hands
on experience has proven to be the most efficient method to train
new users, specially when focusing on how to deal with common
problems. Special attention needs to be given to data retrieval and
manipulation methods in order to allow tracking of data problems
later in the processing chain. Documentation of our own group's
deployment protocols and data handling procedures complemented and
helped with equipment manuals and specifications.

Regularity in systems maintenance is key in keeping them running
within long term deployments. Life expectancy and calibration
deviation for sensors are usually a parameter specified by the
manufacturer. Enviro-Net deployments usually have two maintenance
cycles: one for basic overall system check (and data retrieval for
off-line deployments) and another for complete verification of the
equipment. The basic cycle has intervals ranging from two weeks to
two months, depending on the accessibility of the site and
regularity of visits for other purposes. This task is usually
performed by a member of the local research teams and involves
cleaning the sensors if needed---mostly from dust build-up or
obstructions such as leaves, insect or bird nests, etc--verification
of the general health of the system, and data retrieval, usually the
most relevant part in a basic maintenance cycle. The complete cycle
intervals ranges from 6 to 12 months, and allows detection of a
broader range of problems---e.g., battery charge retention capacity.
This task is usually performed with one more experienced technician.

\subsection{Sensors and Loggers}

Tables~\ref{tab.loggers} and \ref{tab.sensors} list the equipment
used in our deployments. For datalogger systems, shown in
Table~\ref{tab.loggers}, wired and wireless systems are available.
In wired systems all the sensors are connected directly to the data
logger and the communications with it are done mostly through a
physical connection using a cable (serial or USB, for instance) to
connect to a laptop.
For wired deployments, we mostly used {\em Onset Computer Corp.} %
(\url{http://www.onsetcomp.com/}) data loggers; specifically the {\em
HOBO Micro Station}, the {\em HOBO U12 Series} and the {\em HOBO U30
Series} models were employed.

Wireless systems, on the other hand offer different strategies to
eliminate the need for cabled connections.
As an example, the equipment manufactured by {\em Olsonet Communications Corporation} %
(\url{http://www.olsonet.com/}) offers two types of nodes: a collector
and an aggregator. The former is connected to the sensors and is
responsible for wirelessly transmitting the readings to the
aggregator, which works as a centralization point for the data
collection also dubbing as a short term data logger. The aggregator,
however, requires a cable connection for setup or data recovery.
A different strategy is used by the equipment manufactured by {\em MicroStrain, Inc.} %
(\url{http://www.microstrain.com/}), where each {\em
ENV-Link\textsuperscript{\texttrademark}} node works as an
individual data logger, but the connection to these nodes for setup
and data retrieval is done through a wireless connection.

The storage capacity for samples in both types of loggers usually
match the power consumption characteristics to achieve similar
longevity in field deployments. As discussed later in this section a
satellite up-link and a continuous battery recharging capability
(e.g., using solar panels), would allow even longer time spans.
However, since in practice maintenance is necessary long before
these limits are reached, battery and storage lifetimes are not a
limitation for most of these types of equipment.

\begin{table}[!ht]
\begin{center}
{\scriptsize
\makebox[\textwidth]{%
  \begin{tabular}{l l l l l}
  \hline
{\bf Logger Model} & {\bf Connectivity} & {\bf Storage Memory} & {\bf Power (Battery Type)} & {\bf Est. Longevity\textsuperscript{(a)}} \\
\hline
Onset U30 & wired data and setup & 512 KB & Int.~(4.5 or 10 Ah, 4 V) + Solar & solar panel\textsuperscript{(b)}\\
Onset U12 & wired data and setup & 43,000 samples (64 KB) & Int.~(CR-2032 lithium 3 V) & 10--12 months\\
Onset Micro Station & wired data and setup & 512 KB & Int.~(4 x AA 1.5 V) & 10--14 months\\
Olsonet Collector & wireless data / no setup & 256 KB & Int.~(2 x AA 1.5 V) & 4--5 months\\
Olsonet Aggregator & wireless data / wired setup & 2 GB (remov. SD card) & Ext.~(7--12 Ah) + Solar & solar panel\textsuperscript{(b)}\\
Microstrain ENV-Link & wireless data and setup & 360,000 samples & Int. (650 mAh) + Ext. (9 Ah) & 10--14 months\\
\hline
  \end{tabular}
}

  \caption{Dataloggers summary.}
  \label{tab.loggers}
  (a) Estimated longevity with 15 minutes sampling; (b) Dependent on sun light availability.
}
\end{center}
\end{table}

\vspace*{-6truemm}

\begin{table}[!ht]
\centering
  \caption{Sensors summary.}
{\scriptsize
\makebox[\textwidth]{%
  \begin{tabular}{l l l l l}
  \hline
{\bf Sensor Model} & {\bf Variable (Unit)} & {\bf Sensor Type} & {\bf Range} & {\bf Accuracy}\\
\hline
Sensirion SHT-75 & Temp.~(\degree C) & silicon bandgap & $-$40.0--123.8 \degree C & 0.3--1.5 \degree C\\
~ & Rel.~Hum.~(\%) & capacitive humidity & 0--100\% RH & 1.8--4.0\% RH\\
Onset S-THB-M00x & Temp.~(\degree C) & silicon bandgap & $-$40.0--75.0 \degree C & 0.2--0.7 \degree C\\
~ & Rel.~Hum.~(\%) & capacitive humidity & 0--100\% RH & 2.5--4.5\% RH\\
Onset RG3-M & Rainfall (mm/h) & tipping bucket & max 1,270 mm/h & 1.00\%\\
Onset S-LIA-M003 & PAR ($\mu$mol/m2/sec)\textsuperscript{(a)} & photons detector & 0--2,500 $\mu$mol/m2/sec\textsuperscript{(c)} & 5.0\% or 5 $\mu$mol/m2/sec\\
Onset S-LIB-M003 & Solar Radiation (W/m2) & silicon photovoltaic detector & 0--1,280 W/m2\textsuperscript{(d)} & 5.0\% or 10 W/m2\\
Apogee SQ-110 & PAR ($\mu$mol/m2/sec)\textsuperscript{(a)} & photons detector & 0--2,000 $\mu$mol/m2/sec\textsuperscript{(c)} & 5.00\% \\
Apogee SP-110 & Solar Radiation (W/m2) & silicon photovoltaic detector & 0--1,100 W/m2\textsuperscript{(d)} & 5.00\%\\
Decagon ECH2O EC-5 & Soil Moisture (VWC\textsuperscript{(b)}) & 70 MHz capacitance/frequency & 0--100 \% VWC & 1.0--3.0 \% VWC\\
\hline
  \end{tabular}
}
  \label{tab.sensors}
  }

  \scriptsize
(a) Photosynthetically Active Radiation; (b) Volumetric Water
Content; (c) For wavelengths  between 400 and 700 nm; (d)
For wavelengths between 300 and 1,100 nm.
\end{table}

The biggest advantage of wired equipment is reliability, being in
use longer, and tested under many combinations of conditions.
Besides limited spatial coverage, when compared to wireless systems,
the most problematic aspect of this technology is accessibility.
Everything requiring a physical connection between the logger and
the laptop with the control software, having to climb up a tower to
perform tasks as routine as retrieving data is a somewhat serious
limitation. Even using longer cables for the sensors, which still
have a limited maximum length on account attenuation of the electric
signal, towers for higher canopies require climbing to access the
logger.

For environmental monitoring, the major advantage of wireless sensor
systems is the possibility of covering larger areas, without giving
up high spatial and temporal resolution, and at a reasonably low
cost. One low point of the technology is that it is still fairly new
as a commercial product, and still needs some adaptation. Errors in
communication protocols, radio range limitations, power management
related issues, lack of features in the control software packages,
and breaches in weather proofing cases weight in at the cons for
wireless systems. However, our experience shows the technology has
already reached the tipping point to becoming viable for use in long
term, harsh environment deployments.

Commercial availability of {\em wireless sensor networks} (WSN), as
a technology, is still limited. Although the original ideas for
WSN---\emph{i.e.},  large number of general purpose nodes
distributed in very dense deployments, randomly placed, almost
weightless, and disposable---have yet to materialize
\cite{bainsc2008a,rach2008,welsh2010}, wireless technology used in
conjunction with sensory equipment is proving to be invaluable in
monitoring larger areas at the scale of a single ecosystem.

Table~\ref{tab.sensors} lists the main sensors use in our
deployments, which are are well known, commercially available,
inexpensive, and based on established technologies. With the
variables listed, it is possible to extract plenty of derived
information from them, such as vegetation indexes and light
absorption patterns for photosynthesis. In our deployments, we used
solar radiation sensors provided by Onset and {\em Apogee
Instruments, Inc.} (\url{http://www.apogeeinstruments.com/}); air
temperature and relative humidity sensors by {\em Sensirion Inc.}
(\url{http://www.sensirion.com/}) and Onset (The Onset temperature and
relative humidity sensors used are repackaged Sensirion sensors);
and, soil moisture sensors by {\em Decagon Devices, Inc.}
(\url{http://www.decagon.com/}) Lower cost sensors systems usually do
not offer calibration options for the user; they have their
calibration adjusted at the manufacturer (who usually offer
recalibration services).

\subsection{Deployment Configurations}

Within the Enviro-Net Project there are currently two main types of
deployments: phenology towers and understory installations. A
phenology tower uses two solar radiation flux sensors (also called
pyranometers), measuring wavelengths between approximately 300 to
1,100 nm, and two Photosynthetically Active Radiation flux sensors
(or PAR sensors), which measure wavelengths between approximately
400 to 700 nm. Ratios of these measurements can be used to derive
vegetation indexes such as Normalized Difference Vegetation Index
(NDVI) or Enhanced Vegetation Index (EVI)--see, for instance
\cite{hublja1999,jeribr2007,wime2007,rosh2009}. Such indexes can be
used as proxies to monitor vegetation phenology. Understory
deployments are used to monitor the conditions below the canopy
level, and usually cover a larger area.

\begin{figure}[!ht]
  \begin{center}
  \includegraphics[width=\textwidth]{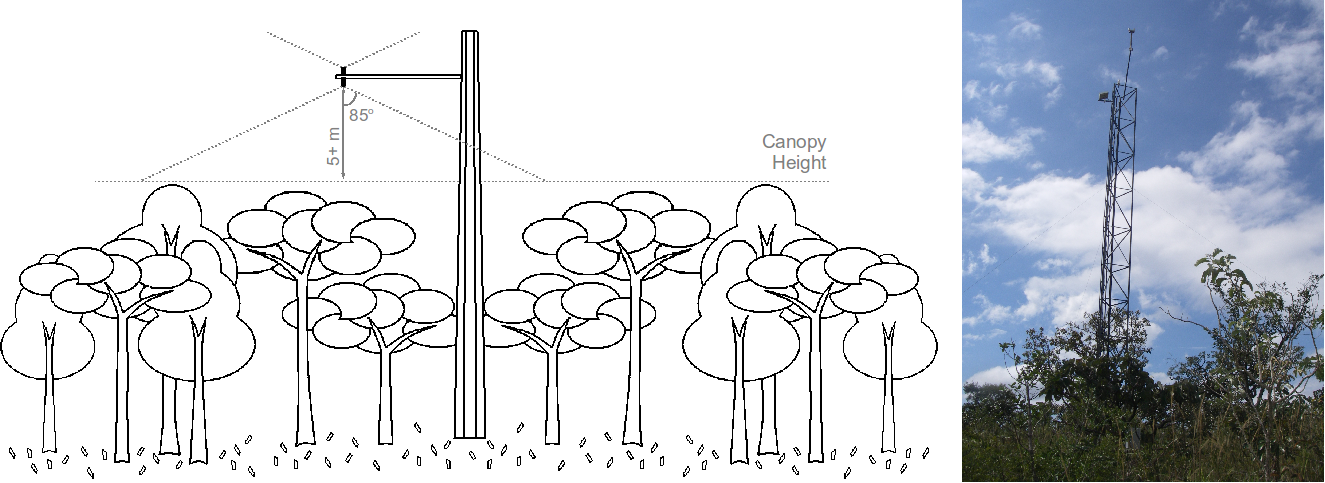}
  \caption{Phenology tower schematics (left) and a tower in Brazil (right).}
  \label{fig.phenology}
  \end{center}
\end{figure}

Figure~\ref{fig.phenology} shows the schematics of a phenology tower
on the left, with two PAR sensors and two pyranometers, one of each
measuring incoming solar radiation and one of each measuring
reflected solar radiation. The right side of the figure is a photo
of one phenology tower installed in Brazil, which raises the sensors
eight meters from the ground, six meters above the canopy.

Most radiation flux sensors have view angles of up to 85$^{\circ}$
from zenith (when oriented up, \emph{i.e.}, measuring incoming
radiation) or nadir (when oriented down), with a uniform
360$^{\circ}$ rotation. With that, the radius that affects the
readings is up to around ten times the distance ($h$) between the
sensor position and the surface being monitored (\emph{i.e.},
$radius = \tan(85^{\circ}) \times h$). For our deployment, we
usually have at least five meters between the top of the canopy and
the sensor measuring the reflected radiation (8 to 15 m
in total), leading to a coverage radius of at least 50 m in the
monitored area.

Obstructions within the range of a sensor interfere with the reading
and might not be easy to identify from the data only---e.g., higher
canopy of adjacent ecosystems or a nearby tower with other
instruments may interfere with sensor measuring incoming radiation.
A sensor measuring radiation reflected from the canopy is more
susceptible to interference---e.g., the positioning of solar panels,
whose reflectiveness greatly affect readings. Large panels should be
positioned outside of the interference radius, while smaller panels
can be positioned at the same height as the sensor for no
interference. Note that it is difficult to position radiation
sensors and solar panels at different orientations, since both
should use the optimal exposure angle to the sun, roughly North, in
southern latitudes, or South, in northern latitudes.

Monitoring the conditions under the canopy level, \emph{i.e.},
understory deployments, allows assessing a different range of
micro-climatic conditions and also soil condition---e.g.,
temperature and moisture levels. Understory deployments are usually
easier to access, and with that, they are useful for validating the
readings observed in a tower and also as a backup for certain
variables in case of sensor malfunction in a tower. Using wireless
systems substantially increases the spatial coverage of understory
deployments with a fraction of the increase in cost and efforts to
retrieve data and maintain the system.

\begin{figure}[htb!]
  \begin{center}
  \includegraphics[width=\textwidth]{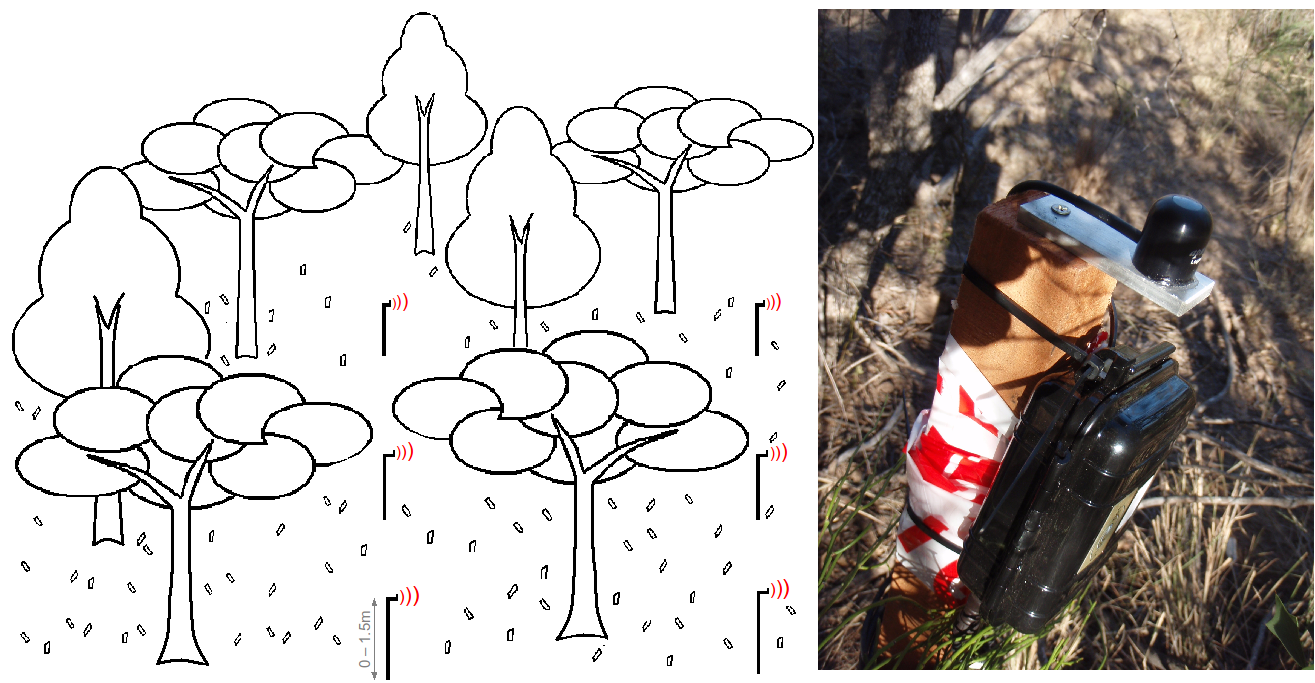}
  \caption{Understory schematics (left) and a node in Argentina (right).}
  \label{fig.understory}
  \end{center}
\end{figure}

Figure~\ref{fig.understory} depicts an example of such a wireless
deployment on its left side. On the right side, it shows a node
deployed in the {\em chaco} ecosystem in Argentina. The height at
which the sensors are installed in this case is also determined by
the canopy's height, usually ranging from right on the ground (e.g.,
for grasslands) to 1.5 m for taller canopies. One example of
application that relies on the spatial coverage and resolution of
understory wireless deployments is deriving Leaf Area Index
(LAI)---see \cite{fuaska1984,wime2007}, for instance. LAI, along
with Plant (PAI) and Wood (WAI) Area Indexes \cite{sakaes2009}, are
important indicators of vegetation productivity, being also used as
a reference for crop growth rates. Combining readings from a
phenology tower with understory readings of absorbed solar radiation
fluxes, it is possible to derive NDVI for the location of each node.
Using NDVI and knowing an appropriate conversion factor,
characteristic to each ecosystem, it is possible to calculate LAI
for each node \cite{wime2007}. This allows the creation of maps of
very high spatial and temporal resolutions for both NDVI and LAI.

\begin{figure}[!ht]
  \begin{center}
  \includegraphics[width=\textwidth]{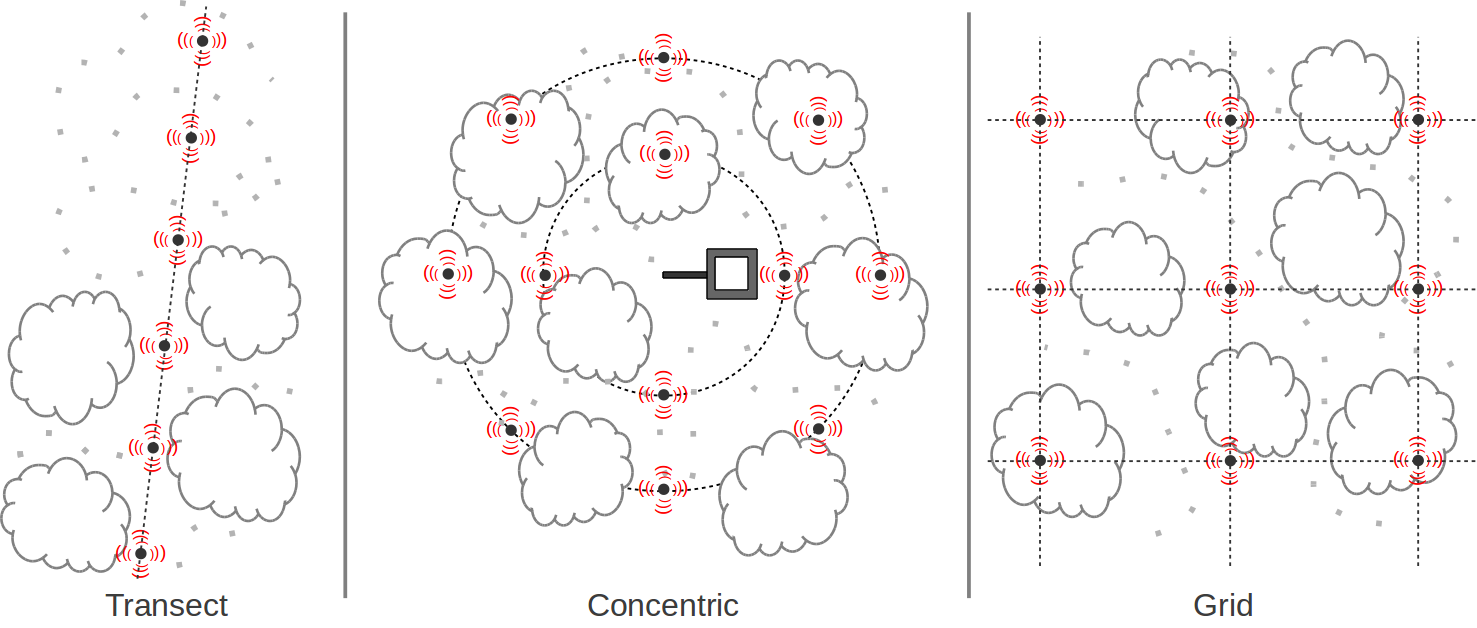}\\
  {\hspace*{-10truemm}(a)\hspace*{50truemm}(b)\hspace*{50truemm}(c)}
  \caption{Deployment strategies: (\textbf{a}) transects; (\textbf{b})
concentric circumferences; and (\textbf{c}) grids.}
  \label{fig.deployments}
  \end{center}
\end{figure}

Having the option of deploying a large number of sensors in a given
area also raises the question of how to distribute these sensors. We
have adopted three different strategies to spatially distribute
nodes and their sensors. Figure~\ref{fig.deployments} illustrates
these strategies. The first approach, shown in the left, is intended
to monitor a linear region along a transect. This is particularly
useful for monitoring transitions between ecosystems or exposition
to different conditions within the same ecosystem. The center of the
figure shows distribution of nodes in concentric circles, which is
sometimes called a ``star'' deployment. This type of deployment is
used mostly to monitor conditions around a point of interest,
usually corresponding to the footprint of phenology or carbon flux
towers, allowing combination of measurements from both deployments.
A third strategy is to deploy nodes in a grid, covering a
potentially larger area of interest. Regularly spaced grids are
useful for uniform monitoring throughout an area. However, irregular
grids can also be useful when special conditions occur within a
region of interest. Examples include part of an area that is also
being monitored by other experiments (e.g., leaf collection for
chlorophyll measurements); or patches affected by fire and
monitoring their recovery is of interest.

\vspace*{-5truemm}
\subsection{Deploying Sensors Systems}

From a logistics perspective, installing tower and understory
systems have fairly different characteristics. Phenology towers
reached up to 15~m in one of our deployments, with 9~m being the
most common height. Selecting the location for installing a tower
that high must take into account the representativeness of the
ecosystem, the impact of building it, and the accessibility to bring
its parts to the site. Another important issue is the uniformity in
the height of the canopy. Too much variation in the tree heights
will lead to scaling problems in the data, an area with taller
vegetation will be contributing significantly less to the readings.
When installing a phenology tower intended to be used in a long term
data collection, the growth of the vegetation should also be taken
into account. Younger ecosystems might grow considerably at
intervals as short as one year, forcing height upgrades to a tower.

The height of the canopy is also a concern for understory
deployments. Ecosystems with lower canopies, such as grasslands,
require that solar radiation flux sensors be positioned almost
adjacent to the ground, while taller canopies allow sensors in a
higher position (0.60 to 0.10 m are common heights). For wireless
deployments, the node is usually installed in a higher position to
improve radio signal range, while the sensors are deployed at the
appropriate heights.

Although it might seem like a trivial task at first, correctly
positioning the sensors should take into consideration a number of
factors. One issue is the creation of unnatural sources of shade
(e.g., from the pole where the node sits) into the sensor. For
deployments in the northern (southern) hemisphere, positioning
radiation flux sensors South (North) of obstructions avoids this
issue. Air temperature and relative humidity sensors are also
affected by their positioning. Besides being hosted at solar
radiation shields and being positioned as to allow for air
circulation, they should also keep some distance from radiation
absorbing materials. Most of the weatherproofing cases, for
instance, absorb non-negligible quantities of solar radiation. We
had cases of temperature deviations of up to 20 $^{\circ}$C because
of a dark weatherproof case.

One crucial aspect to sensor systems deployments in tropical
ecosystems is the exposure to constantly high relative humidity.
Values between 90\%--100\% are common in these environments.
Combined with high temperatures, this condition transformed many
weatherproof casings into humidity traps. The main problem was
actually the difference of internal and external pressure in the
cases. That made previously air tight cases absorb humidity while
balancing the pressure, exposing the internal circuits and
connectors. Both for loggers and sensors, even cases designed and
tested to work underwater were susceptible to this problem. Adopting
pressure relief valves significantly attenuated the problem, even
though sometimes they can get clogged with dirt and stop working.
Another adopted practice that also helped reduce this problem was to
use silicone-based adhesives to seal borders and openings, around
sensor cables and also around the sensors themselves.

For wireless sensor systems, testing the range of the radio system
at the actual deployment site is essential. Vegetation distribution
and terrain contours are difficult to predict beforehand and have a
significant impact in the radio range. Two major aspects have shown
to be of particular relevance when conducting this kind of test.
Firstly, if the type of batteries used decrease the voltage offered
to the system with time, the tests should not be conducted with new
batteries. A more accurate test of radio range is achieved using
more realistic battery levels---e.g., levels of battery similar to
when a deployment running for more than half of the expected battery
life. In case of rechargeable batteries, the charge level used
should be the average level the batteries would have when going
without charge for the maximum foreseeable period. For tropical dry
forests the maximum period without non-negligible sun light exposure
for charging batteries through solar panels is around two days. It
is worth of note that regular alkaline (zinc and manganese dioxide)
batteries, widely adopted to power nodes in sensor networks, do
change their voltages depending on their level of charge.

The second aspect interfering with radio range is related to the
vegetation density, particularly to changes in density throughout
the seasons. Radio range is greatly affected by branches and leaves
in the line of sight of the signal. Ranges of up to 300 m in a level
and open field can be reduced to as little as 15 m (a factor of 20
reduction) simply by having a somewhat closed vegetation. In
particular, radio frequencies at 2.4 GHz are severely attenuated by
trees and leafs.
This frequency is adopted by several wireless sensor systems, including the ZigBee Alliance %
(\url{http://www.zigbee.org/}) communications protocols (based on the
IEEE 802.15.4 Wireless Personal Area Network standard), widely used
in these systems. Furthermore, it is very usual for deployment
campaigns to take place in the dry season, when rainfall is less of
a concern for the schedule in deployment plans. However, foliage of
deciduous vegetation can be at much lower levels than it will be in
the wet season, which can significantly affect the range of radio
signal. There is no definite solution to address the vegetation
changes, since simulating the conditions of a different season is
difficult. Monitoring the overall network health, which can be done
in its simplest form by detecting gaps in the data, and
repositioning nodes when necessary has been the best measure to
address this issue in new ecosystems. These, in turn, serve as a
reference for future deployments in similar ecosystems.

Seasonal change also can have an unexpected impact in the visibility
of nodes and sensors. When installing sensors in the dry season,
there are few obstructions and less color variability on the
landscape. This makes visibility reasonably good. However, areas
that significantly change their vegetation coverage or areas that
have dense vegetation can become quite challenging from the point of
view of visibility in the wet season. Using colorful markers---red
or yellow ribbons or paint are effective for this---can save a lot
of time when trying to find nodes and sensor that have been deployed
for a while. Not relying solely on the GPS to locate small pieces of
equipment such as individual nodes and sensors can be the difference
between returning to base camp before or after sunset. One aspect of
using such markers that was not taken into account in this work is
the increased attractiveness color makers might exert on animals
(particularly insects).

\subsection{Retrieving Data}

Data from ground-based sensor systems can be retrieved either {\em
in-situ} or remotely. The former involves expeditions to the
deployment sites, which can be very expensive. However, if the site
is already being visited in a regular basis for other reasons
(collecting leaf samples, for instance), this might become more
feasible. Most of our current deployments are working in this
scenario. This has proven to be quite an advantage from the
perspective of maintenance of untested systems, allowing early
detection of problems with equipment. With equipment proven to work
well, using a remote solution is probably more cost and time
effective.

Collecting data remotely might be achieved in a number of ways. One
possibility is using a dedicated long range wireless communication
system---e.g., by using a WiFi connection with repeaters---to
transmit data at regular intervals to a computer installed in a
location with permanent power supply. If there is also Internet
connectivity, the data can be forwarded to on-line permanent
archival systems. This alternative usually has a significant
overhead of maintaining the local computer and the long range
communications system running.

Another alternative is to use cellular networks with data
capabilities. Although cellular coverage is not good in more remote
areas, some regions have enough connectivity to allow data
transmission in a fairly regular fashion. Using higher gain antennas
improves signal reception, but the system must be prepared to go
through reasonably long periods with no connectivity, preserving all
data for delayed transmission. Since an actual Internet connection
is provided with a cellular connection, the data can be transmitted
directly to on-line archival servers.

A third type of remote data retrieval solution involves using a
satellite up-link. This approach is also subject to communications
failure (e.g., if there is too much cloud coverage). The
connectivity provided here usually is not to the Internet, but
connectivity to a service provider that receive the data from the
satellite. This provider in turn makes the data accessible, often
offering automated ways of retrieving the data from their on-line
servers. In our case, systems that have proven to work consistently
well have been equipped with a satellite transmitter.

Remote connectivity allows not only automated data retrieval, but
also some level remote operation of the equipment. Options of
stopping and starting the logger, setting sampling and storage rates
are often available. In a few cases, it might be interesting to be
able to set other parameters remotely, particularly with wireless
systems. Research projects have explored configuring deployments
remotely \cite{stda2009}, even reprogramming loggers and collection
nodes in some cases \cite{bainsc2008a,abhasa2007}. This level of
flexibility in remote deployment configuration, however, is not yet
commercially available.

\section{Data Pre-Processing and Cleaning}
\label{ingestion}

When considering the volume of data generated by current sensor
systems, automation of data management related tasks within a proper
computational infrastructure is of paramount importance
\cite{teillet2010,rugral2009,mutesz2006,hama2006,arfako2004,esmibo2003}.
However, actual datasets generated by sensor systems might present a
variety of problems and exceptions, which are often difficult to
foresee. This is a severe drawback in attempts to automate the first
data management phase: ingestion of data into any computational data
management system. This sort of problems are often dismissed as
being ``implementation details'', but their implications can
actually affect data quality parameters and models to store and
distribute the data. In higher end (expensive) and/or homogeneous
equipment this sort of problems are usually easier to tackle.
However, in a setting like ours, using equipment from different
manufactures, in a highly distributed effort, with an aim at low
limits for equipment and maintenance costs, these issues are
fairly~commonplace.

The implementation of solutions for problems with raw datasets are
usually carried out within a {\em data pre-processing} (or {\em data
cleaning}) phase. Although these terms usually encompass explicit
data quality verification or removal of erroneous readings (e.g.,
values outside the scale measured by a sensor), this section only
considers problems that actually prevent (or are difficult to trace
after) the ingestion of the data into a data management system. When
compared to classification scales usually adopted in describing
Earth observation data products, after the corrections in this
section, the dataset should treated as ``raw'' data, or, as being at
Data Processing Level 2 in the National Research Council (NRC)
Committee on Data Management and Computation (CODMAC) \cite{pds2009}
classification, or to Data Processing Level 0 used by NASA
 (\url{http://science.nasa.gov/earth-science/earth-science-data/data-processing-levels-for-eosdis-data-products/}).
The next paragraph discuss the problems we had to handle when
preparing our datasets.

\subsection{Synchronization}

Keeping correct temporal information for timestamping readings from
distributed sensors can be really challenging, not to mention
correcting time deviations after recording the data
\cite{gumusz2009}. Time synchronization is an issue both at single
deployment, with multiple collectors and/or loggers, and across
deployments. Within a deployment, hardware imprecision and
heterogeneous initial synchronization methods are the two main
causes of synchronization problems. Time keeping in electronic
equipment is based on crystal oscillators, which can deviate from
their standard frequency with environmental conditions, especially
temperature. This causes the time measurements to deviate as well,
and affects almost all types of data logging equipments. In this
case, the error is proportional to the sampling rate, which for
applications such as seismology, with high sampling rates are, these
errors are quite significant. For long term environmental
monitoring, this can also be a problem. One solution is to have an
accurate reference time keeping and a mechanism to keep the
synchronicity among loggers. Possible solutions include having more
precise equipment kept at a less exposed location or using GPS time
as references. A few wireless communication protocols have time
synchronization features embedded within their message exchanging
mechanism \cite{subuks2005}.

When dynamic time synchronization against a reliable reference is
not feasible, the initial synchronization method is the basis for
all time information within a deployment. This is the most common
scenario for our current deployments, with the usual mechanism for
synchronization being based on the time information from the
computer with the control software used to start a deployment.
Therefore, the time information in that computer should be
synchronized (e.g., by using Network Time Protocol, IETF RFC 5905
(\url{http://tools.ietf.org/html/rfc5905})).

Data comparison from different deployments at small temporal
resolutions must take into account potential synchronization errors.
However, since sampling rates are commonly higher than desired
temporal resolution, most data analysis is done with aggregated data
instead of the entire dataset, which attenuates the effects of the
time synchronization related errors, particularly when looking at
hourly or even daily averages.

Similar to other reports \cite{gumusz2009}, we also experienced
power source related synchronization problems. Time measurement in
some logging equipments can be affected by power outages or low
voltages from the power source. Some types of equipment use the main
power to keep time measurement running and, although time
measurement usually requires very little power, if the supply is
interrupted, the equipment's clock gets reset.

Current data logging equipment and control software offer poor
support to address time synchronization problems. Many of them don't
even let the user see what is the current time in logging system to
manually check for time drifts. But this is evolving in control
software for wireless systems, since these suffer more noticeably
from time related problems.

\subsection{Time Zones}

When dealing with deployments that are geographically distributed
throughout various timezones, establishing the correct local time
can become an issue. Once again, relying on a computer's time as a
reference to timestamp the readings is a major cause of errors.
Different versions of operating systems have different levels of
automation regarding time zones and daylight saving time
configurations, often allowing users to change these manually.
Therefore, besides having the correct time on the computer, as
already discussed, wrong configurations of time zone and changing
configurations for daylight saving times can also lead to
inconsistencies such as: having data for a single deployment
timestamped with different daylight saving times, or difficulties
determining which is the correct local time when comparing data for
deployments in different time zones.

For our deployments, when issuing field laptops, time configurations
always adopt the local standard time for the site, disabling
automatic changes to daylight saving time. However, even rugged
field laptops fail, and temporary misconfigured replacements can be
used. Or, an even less elaborate problem, which happens often, new
users get confused by seemingly ``wrong'' time settings and change
the time configurations.

It is possible, however, to check time zone and daylight saving
times against sun time. This is done by comparing several days of
sunrise time from data collected by solar radiation sensors to
expected sunrise times for the location. This method is not accurate
enough for correcting for hardware time drift, for instance, but is
good enough to correct for one or more hours shift in the
timestamps. This verification is performed on all of our datasets
before ingesting them into our data management system.

\subsection{Data Format Variation}
\label{dataform}

One burdensome problem of dealing with data from different types of
equipment is handling changing data formats or a variety of possible
formatting errors.

The first of such types of problems to be addressed are changes in
the data format made by the equipment manufacturer. A considerable
amount of format changes from manufacturers are not documented
adequately with new versions or software updates. Unfortunately,
this type of problem needs to be addressed case-by-case.

One problem that was surprising to us is that some types of failures
in the sensors themselves can generate errors in the data format by,
for instance, changing the number of data columns in a record. As an
example, this could make a record that should have three data
columns (e.g., readings on temperature, humidity and solar
radiation) actually have extra or missing columns. Similar effects
can be caused by connector designed to be generic and support
different sensors: a sensor behaving in some unexpected way may
cause the data collection node or the logger to perform incorrect
conversions or even generate software errors that will affect the
data format. In wireless equipment, we have also seen the data
format being drastically changed by problems in the transmission of
the data. In the presence of radio interference, usually created by
the operation of higher powered wireless equipment in proximity of
the deployment, the data transmission gets compromised, generating
errors in both the values of the readings and the structure of a
record.

All these types of errors can cause failures in the ingestion
software or, worse, have errors introduced in the data ingestion
process without any warnings. Our solution to this was to make
ingestion software monitor for format changes and generate
informative error messages, allowing identifying problems before
ingestion.

\subsection{Provenance}

Given the tailored nature of data pre-processing steps described in
this section, it is difficult to keep standardized provenance
information and even harder to automate collection of this type of
information, as also made evident in \cite{waboma2007}. In our
current data management solution, simple free text description
fields are used to keep track of pre-processing steps and choices.
Nonetheless, with a flexible metadata specification tool, such as
the one created for our partner project GeoChronos
(\url{http://www.geochronos.org/})~\cite{cukisi2010}, it would be
possible to add specific fields to document evolving aspects of the
pre-processing~steps.

Although difficult to obtain and maintain, documentation of the
pre-processing steps are important to identify not only errors in
the pre-processing itself, but also problems with the deployments.
For instance, the appearance of too many erroneous records from a
wireless data collection node are potential evidence of problems
with the sensors, the sensor connections, the node hardware or radio
interference sources in the surroundings. The latter problem might
even indicate affected readings from other nodes that would
otherwise go unnoticed.

\section{Web-Based Data Analysis}
\label{web}

A resourceful and easy to use data management system is the last
piece of our solution for large scale {\em in-situ} monitoring. The
pre-processing step presented in the previous section allows the
data to be ingested into the system, being stored in an integrated
representation. Then users can interact with the system having
access to data filtering, aggregation and other more specialized
processing operations. Data visualization and retrieval are offered
for data at all processing levels after pre-processing, providing a
flexible mechanism for analyzing the data within the system or using
other tools with the data already narrowed down to the parts of
interest. This section discusses these issues, also considering data
quality and user interaction aspects.

\subsection{Uploading Data}
\label{up}

The task of data ingestion can be automated for datasets that
require pre-processing steps known beforehand. Automated data
ingestion methods are particularly useful with deployments that have
automated data retrieval, as is the case for data retrieval using a
satellite up-link and an the respective Internet service for getting
the data. However, new datasets or datasets that needed specialized
pre-processing before getting ready to be ingested need a flexible
mechanism to map available data to the integrated representation of
the data in the system. Properly handling errors and exceptions in
the data ingestion processing is necessary from both user experience
and data quality perspectives. Automated data ingestion needs timely
error generation so the user responsible for the deployment is kept
informed and and can take corrective action. Informative
descriptions of errors helps the user identify and diagnose the
error causes, which is particularly important for manual ingestion
of data that was pre-processed in any non-standard way.

Another aspect to be considered is the documentation of the process
for every dataset upload. Metadata regarding date and time, data
source, user, pre-processing options, among others, help identify
sources of error such as faulty time related correction or
application of outdated pre-processing methods. Most of these
metadata can be collected automatically by the system, which
unburdens the user and prevents missing information from less
thorough users.

Only with an integrated data representation model it is possible to
offer a common user interface, types of filters, aggregation options
and any other operation for manipulation of data. Data from
different instruments, deployment configurations, retrieval
strategies, \emph{etc.}, need to be stored uniformly so all the
system's features are available for all datasets.

\subsection{Interactive Filters and Operators}
\label{filters}

\begin{figure}[!ht]
  \begin{center}
  \includegraphics[width=\textwidth]{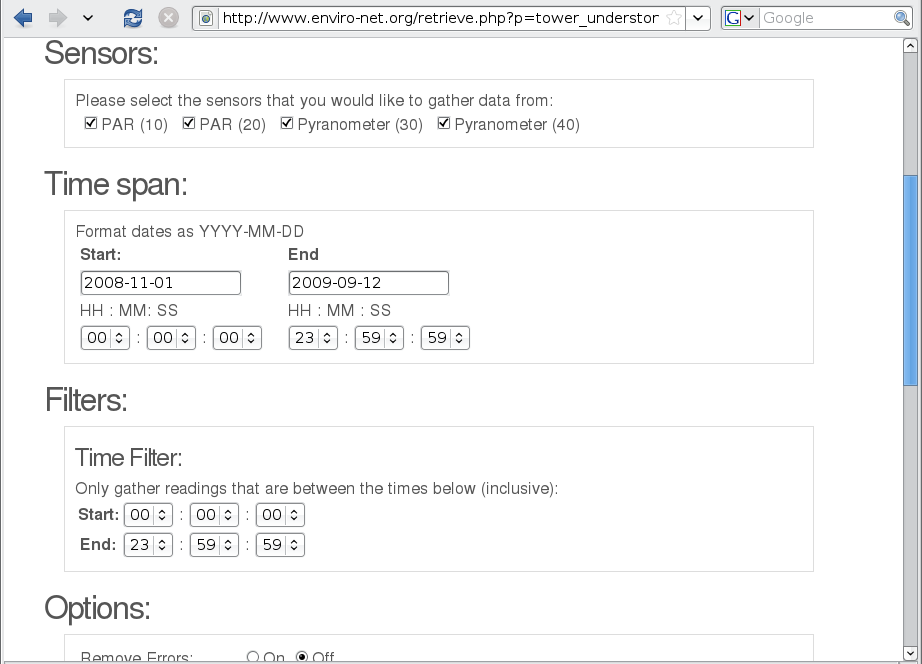}
  \caption{Data retrieval options.}
  \label{fig.screen.retrieve}
  \end{center}
\end{figure}

Having the data uniformly stored in a repository, the users can
start tailoring datasets to their needs. The most basic
functionality to allow this tailoring is being able to apply filters
(e.g., only data within a range of values or with low error rates)
and aggregation operators (e.g., showing daily or monthly values) to
the datasets. Without adequate computational support, many
researchers spend days to weeks in this trivial task.
Figure~\ref{fig.screen.retrieve} shows our interface for a few of
these filters to achieve the target data, from the top: which
sensors to include, which time span to consider, and which times of
the day are of interest. The screen shown is to extract and download
a dataset to be used with other tools. Several other options are
also available, including filtering out errors, showing raw values
(e.g., of voltages, electrical current, or unconverted pulse
counts), file and content formats, including data derived from the
sensor readings, among others.

Offering quick and easy access to the (corrected and quality
checked) sensor readings from a deployment is one of the most
essential features of our solution. However, also having data that
can be derived from from these readings as easily accessible is what
shows the actual potential for data management systems like ours.
Our current implementation offers the automatic calculation of
vegetation indexes, NDVI and EVI, from solar radiation flux sensors
using different methodologies
\cite{jeribr2007,wime2007,hublja1999,rosh2009}. Other products are
currently being integrated into Enviro-Net, including LAI, {\em
Vapour Pressure Deficit} (VPD), spatial distribution for {\em
Fraction of Absorbed Photosynthetically Active Radiation} (fAPAR).

\subsection{Visualization}
\label{vis}

\begin{figure}[htb!]
  \begin{center}
  \includegraphics[width=\textwidth]{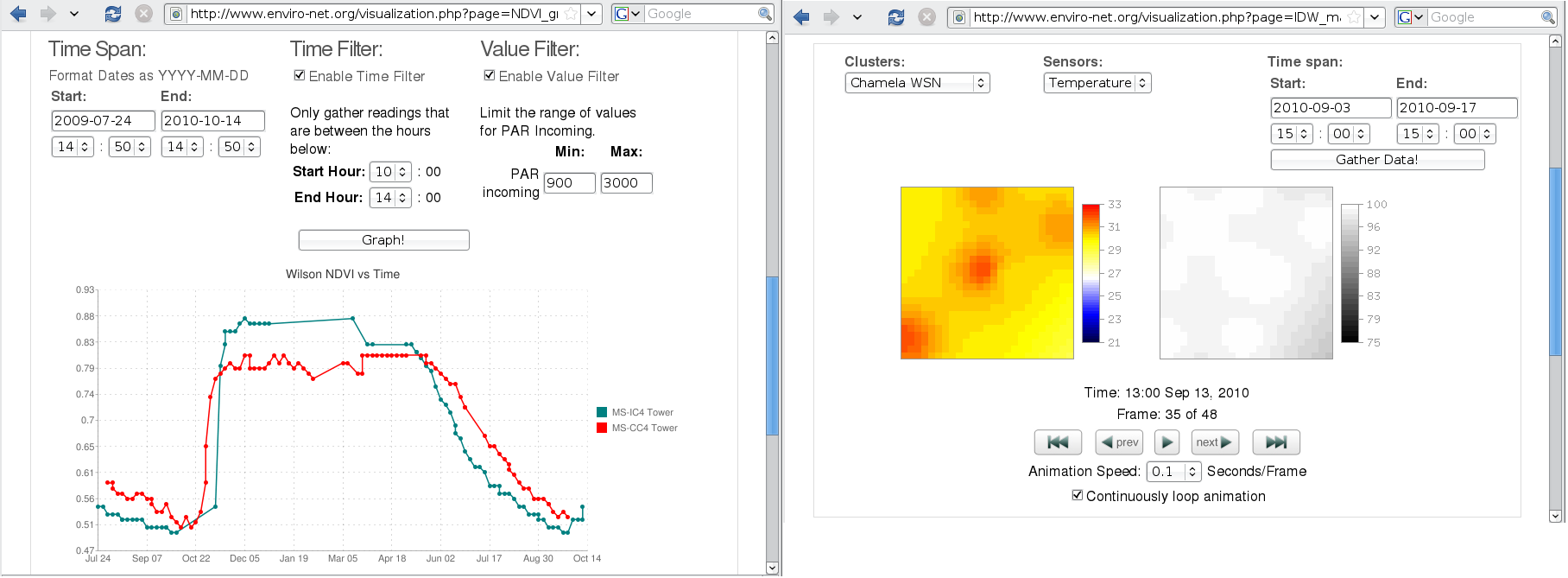}
  \caption{Visualization of derived NDVI (left) and spatial distribution
of temperature and its reliability (right).}
  \label{fig.screen.ndvi}
  \end{center}
\end{figure}

After tailoring a dataset to specific goals, adequate visualization
tools allow easier understanding of events and trends within the
monitored areas. The most basic visual tool is graphing the sensor
readings of different variables, allowing visual comparison and
insight on the measurements in one deployment. However, two features
in our web system proved to be invaluable: graphing of datasets that
went trough transformations (filtering, aggregation and derived
data) and graphing across deployments. These graphing options are
depicted on the left side of Figure~\ref{fig.screen.ndvi}, which
shows derived NDVI (using the methodology in \cite{wime2007}) for
two different deployments in the Mata Seca State Park, in Brazil,
within a specific time span, using only readings close to midday
(between 10:00 AM and 2:00 PM local time), filtering out seemingly
cloudy days (\emph{i.e.}, including only data records when the
measured incoming PAR is more than 900 microeinsteins per second per
square meter---$\mu$E/m$^2$/s), and aggregating the data in daily
averages.

The right side of Figure~\ref{fig.screen.ndvi} shows another type of
visualization strategy based on the spatial distribution of the
readings. The graph on the left site uses a color scale to represent
variation temperature across an area covered by 12 temperature
sensors in the \emph{Chamela Reserve} in Mexico. The graph on
the right shows the coverage of the installed sensors (indicating
the reliability of the scale), highlighting sensor failures when
these occur. Within the specified time span, the system generated a
sequence of images which are animated using the controls at the
bottom to show the evolution of the temperature and reliability
distributions through time. This is a useful tool to observe cyclic
(e.g., diurnal or seasonal) changes in the monitored areas.

\section{Concluding Remarks}
\label{conclusion}

This paper presented the Enviro-Net Project, which addresses a
variety of issues related to {\em in-situ} (or ground-based)
monitoring of ecosystems, from the deployment of sensors to the
delivery of processed data products. A combination of factors make
this project unique: (i) acquisition of data at ecosystem level with
high spatial and temporal resolutions; (ii) long term, ground-based
monitoring; (iii) use of heterogeneous, commercially available, and
inexpensive equipment, including wireless sensing technologies; and
(iv) integrated data management solution, with a Web-based user
interaction with data products.

This scenario, which is increasingly being adopted by other research
projects, is described in detail in the paper, discussing lessons
learned and pointing out aspects that require attention and could go
unnoticed before deployment efforts are well underway. The paper
examines not only technical issues of deploying ground-based sensor
systems, but also the logistics behind execution and maintenance of
deployments, issues related to data retrieval, verification and
quality, and publication of data products. The paper discussed
evidence that this kind of research was needed, integrating solution
to from a number of research efforts and offering a real solution
in-situ long term environmental monitoring at high resolution
temporal and spatial.

Current efforts include: improving our deployment protocols to deal
with arising problems and simplifying the maintenance related tasks;
extending our data management system in order to handle larger
amounts of data; and adding new data manipulation operations to
offer more derived data products. As future work, we intend to focus
on data provenance visualization issues, to improve understanding of
how data products were generated and allowing automation of
reproducibility. Another aspect to be explored  in future releases
of our data management system is the integration of remote sensing
data (from satellite and airborne instruments) into our common
interface \cite{muko2009}, allowing analysis and comparison of these
types of data with ground-based data. We also plan to work on
implementing programmatic interfaces to allow software-based access
to our data by, for instance, using Open Geospatial Consortium
protocols. Lastly, we have plans to include monitoring of equipment
life expectancy, particularly of sensors and wireless collector node
equipment, in order to create better models for maintenance of long
term deployments---by, for instance, increasing the precision of
required replacement rates for equipment.

\section*{Acknowledgements}

The Enviro-Net Project is funded by the Canadian Foundation for
Innovation and the Inter-American Institute for Global Change
Research (IAI) CRN II \# 021 which is supported by the National
Science Foundation (Grant GEO-0452325). It is also partially
supported by Cybera and Canarie (through the GeoChronos Project), as
well as the National Science and Engineering Research Council of
Canada (NSERC) Discovery Grant Program. The authors acknowledge and
appreciate the contributions to the Enviro-Net Project, received in
various forms, by numerous members of the local and remote research
groups, in particular the students and staff at the University of
Alberta, and our research partners at Federal University of Minas
Gerais (UFMG), State University of Montes Claros (UNIMONTES), and
State University of S\~ao Paulo (UNESP), in Brazil; Autonomous
National University of Mexico (UNAM), in Mexico; Technology
Institute of Costa Rica (ITCR), in Costa Rica; University of Buenos
Aires (UBA), and National University of San Luis (UNSL), in
Argentina.

\bibliographystyle{plain}
\makeatletter
\renewcommand\@biblabel[1]{#1. }
\makeatother

\end{document}